\documentclass[11pt,a4paper]{article}
\usepackage{jheppub}
\pdfoutput=1

\usepackage{
graphicx,
subfig,
rotate,
color,
amsmath,
amssymb,
latexsym,
relsize,
dcolumn,
bm,
epsfig
}

\begin{document}

\newcommand{\yes}{\checkmark}
\newcommand{\nil}{$\times$}
\newcommand{\afb}{${\cal A}^t_{FB}$}
\hyphenation{FeynCalc}
\hyphenation{MadGraph}
\hyphenation{PYTHIA}
\def\lsim{\mathrel {\vcenter {\baselineskip 0pt \kern 0pt
    \hbox{$<$} \kern 0pt \hbox{$\sim$} }}}
\def\gsim{\mathrel {\vcenter {\baselineskip 0pt \kern 0pt
    \hbox{$>$} \kern 0pt \hbox{$\sim$} }}}

\catcode`\@=11
\def\sla@#1#2#3#4#5{{%
 \setbox\z@\hbox{$\m@th#4#5$}%
 \setbox\tw@\hbox{$\m@th#4#1$}%
 \dimen4\wd\ifdim\wd\z@<\wd\tw@\tw@\else\z@\fi
 \dimen@\ht\tw@
 \advance\dimen@-\dp\tw@ \advance\dimen@-\ht\z@
 \advance\dimen@\dp\z@
 \divide\dimen@\tw@ \advance\dimen@-#3\ht\tw@
 \advance\dimen@-#3\dp\tw@ \dimen@ii#2\wd\z@
 \raise-\dimen@\hbox to\dimen4{%
 \hss\kern\dimen@ii\box\tw@\kern-\dimen@ii\hss}%
 \llap{\hbox to\dimen4{\hss\box\z@\hss}}}}

\def\slashed#1{%
 \expandafter\ifx\csname sla@\string#1\endcsname\relax
{\mathpalette{\sla@/00}{#1}}
\fi}
\def\declareslashed#1#2#3#4#5{%
 \expandafter\def\csname sla@\string#5\endcsname{%
#1{\mathpalette{\sla@{#2}{#3}{#4}}{#5}}}}
 \catcode`\@=12
\declareslashed{}{/}{.08}{0}{D}
 \declareslashed{}{/}{.1}{0}{A}
 \declareslashed{}{/}{0}{-.05}{k}
 \declareslashed{}{/}{.1}{0}{\partial}
 \declareslashed{}{\not}{-.6}{0}{f}

%
%
%
%
\title{Same-sign Tops: A Powerful Diagnostic Test for Models of New 
Physics}
\author[a]{David Atwood,}
\author[b]{Sudhir Kumar Gupta,}
\author[c]{and Amarjit Soni}

\affiliation[a]{Dept. of Physics and Astronomy, Iowa State University,\\
 Ames, IA 50011, USA}
\affiliation[b]{ARC Centre of Excellence for Particle Physics at the Terascale,
School of Physics, Monash University,\\
Melbourne, Victoria 3800 Australia}
\affiliation[c]{ Theory Group, Brookhaven National Laboratory, \\
Upton, NY, 11973, USA}

\emailAdd{atwood@iastate.edu}
\emailAdd{Sudhir.Gupta@monash.edu}
\emailAdd{adlersoni@gmail.com}

\abstract
{

We study the connection between the same sign top (SST) and the top quark forward-backward asymmetry (${\cal 
A}^t_{FB}$). We find that a large class of new physics (NP) models that have been proposed to account for the 
${\cal A}^t_{FB}$ lead to SST quark production rate much larger than the observed rate at the LHC and 
consequently are severely constrained or ruled out. Our model independent, general, operator analysis shows 
that none of the tree-level self-conjugate flavor-changing operators are able to explain ${\cal A}^t_{FB}$ and 
simultaneously remain consistent with the same-sign top-quark production constraints from the LHC data.
}


\maketitle

\section{\label{sst:intro}Introduction}

The forward-backward asymmetry of the top quark, ${\cal A}^t_{FB}$ 
as measured by the two Fermilab Tevatron 
experiments~\cite{Abazov:2011rq, Aaltonen:2011kc} shows $\sim 2.5 \sigma$ 
deviation from the Standard model (SM) estimates~\cite{nlofb, nnllfb} and 
thus hints at some form of new physics in the top sector. On the other 
hand the measured top-pair production cross-section appears to be 
consistent with the SM prediction within errors. This creates tension in new physics models trying to 
explain the experimentally observed ${\cal A}^t_{FB}$ as the new 
interaction tends to modify the top-pair production cross-section as 
well.

Over the years, several speculations~\cite{afbref, afbmi} have been made to 
explain the observed value of ${\cal A}^t_{FB}$. Some key review papers can be found in the Refs.~\cite{afbmi}.
Axi-gluons, 
color-octet, color-sextet, models with an extra vector boson(s) and/or 
scalar(s) etc. have been introduced to explain this effect. Depending 
upon whether the additional contribution to the ${\cal A}^t_{FB}$ is due 
to a s-channel exchange or due to a t-channel exchange, these 
suggestions fall into two broader categories. The former category of 
models require processes where the exchanged particle has flavor 
diagonal couplings to the SM-quarks. This is severely constrained from 
the direct searches on non-observation of any weakly coupling 
non-standard under-TeV resonance at the Tevatron or the LHC. The latter 
possibility requires processes in which an up- or charm-quark can 
transit into the top-quark or vice versa. Interestingly such 
(self-conjugate flavor-violating) interactions can also contribute in the production of 
the same-sign top pairs at the hadron collider and for this reason we focus 
on the same sign top pair production in this paper.

If these  flavor-violating couplings of the top-quark 
are introduced to explain the  ${\cal A}^t_{FB}$ then,  at the LHC,  pairs of same-sign top quarks can be produced 
via t-(and u-) channel exchange of  scalar, vector or even  tensor 
particles through the parton level process $u u \longrightarrow t t$ 
or ${\bar u} {\bar u} \longrightarrow {\bar t}{\bar t}$. Same-sign 
top-pair production  process is thus a  very useful ``smoking-gun`` signature especially since 
it has 
little SM background and can be observed in the form of a pair of 
same-sign leptons in association with  a b-jet pair due to semileptonic 
decays of the produced top-quarks. Recall that in  models such as supersymmetry,  
due to the Majorana nature of  the gluino,  one expects to observe lots of such 
events when a pair of gluinos are produced and decay into tops in turn. 
The other known class of models where same-sign top production can be 
abundant are those where a boson with charge ($\pm 4/3$) 
couples with tops and thus can give rise to SST-pair via s-channel 
resonance. It is to be noted that SUSY same-sign top events differ from 
our case in the sense that,   SUSY events  carries largee amount of missing 
tranverse energy due to missing lightest stable superparticles, and the 
event rates for both $tt$ and ${\bar t} {\bar t}$ are the same, 
because top and anti-top decays of gluinos have the same rate. The other 
case mentioned is also  different from ours in the sense that there we explicitly 
expect to observe a charge-$\pm 4/3$ resonance.

Note also that: (1) this production process has very little SM 
background as we briefly discuss in section \ref{sst:anal}, and, (2) the 
top quark is a self-analyser of its spin, several interesting 
observables can be constructed in order to test the models.

Thus the interplay among ${\cal A}^t_{FB}$, $\sigma_{t\bar{t}}$ and the 
$\sigma_{tt}$ seems very useful to explore. These three-fold 
experimental measurements should allow us to set up a strategy to single 
out the true new physics operators responsible for explaining the 
measured ${\cal A}^t_{FB}$.

Motivated with this, in the current paper we will establish some 
remedial measures for the ${\cal A}^t_{FB}$ by exploiting all possible 
set of self-conjugate flavor-changing operators of the top-quark which could be 
responsible for the observed deviation at the Tevatron.

Before closing this section, we would like to point out that because LHC is a symmetric $pp$ machine, ${\cal A}^t_{FB}$ 
can not be defined at the LHC. However an analogus (parity-violating) observable to the ${\cal A}^t_{FB}$, called 
charge-asymmetry of the top-quark, $A_C$, can be still defined by the difference of number of $t{\bar t}$ in forward and 
backward rapidity region. Now since, at the LHC $t\bar t$ prodcution is dominated by the gluon-fusion, the $A_C$ is 
expected to be rather small, and therefore, it may not be an useful observable to study the flavor-violating operators 
under consideration and it is not our focus in this paper.

Organisation of the paper is as follows: In section \ref{sst:opr} we will provide a detailed account 
of the most general form of operators that could contribute to the ${\cal A}^t_{FB}$. In section \ref{sst:corr} we will discuss the correlation of the ${\cal A}^t_{FB}$ and the same-sign top-pair cross-section. We provide a detailed numerical analysis for each of the operators in light of the experimental measurements in section \ref{sst:anal}. We conclude with the findings of this work in section \ref{sst:concl}.

\section{\label{sst:opr}The General New Physics Operators}

As reported in Ref.~\cite{Abazov:2011rq, Aaltonen:2011kc}, the 
independent measurements in the $l + j$, where $l = e^\pm, \mu^\pm$ and 
$j$ is a jet, detection mode by the two experiments at the Fermilab 
Tevatron suggest ${\cal A}^t_{FB}$ to be ${\cal A}^t_{FB} ({~\rm\tt 
D\O}) = 0.19 \pm 0.065$, and, ${\cal A}^t_{FB} ({~\rm\tt CDF}) = 0.158 
\pm 0.074$ which are clearly consistent with each other within 
experimental errors yielding a weighted average of $0.176\pm 0.05$. This 
is somewhat higher than the the SM predictions of $0.058\pm0.009$ and 
$0.072 ^{+ 0.011}_{- 0.007}$ at the {\tt NLO}~\cite{nlofb} and {\tt NLO 
+ NNLL}~\cite{nnllfb} levels respectively (note that at the tree level 
the asymmetry is identically zero). Thus the experimental results deviate 
by about $1.7 \sigma$ from the SM. Since the asymmetry in the high mass 
region, ${\cal A}^{t, high}_{FB}$, tends to be a stronger discriminant of models, it may be useful 
to mention that ${\cal A}^{t}_{FB}$ in the high $m_{tt}$ region is 
experimentally (see Table 1) found to be $0.188 \pm 0.043$, whereas SM 
predicts $0.088\pm 0.013$~\cite{Campbell:1999ah} which is again off by $\sim 2.3 \sigma$.

\begin{table}
\centering
{
\begin{tabular}{|c|c|c|}\hline\hline
{\bf Observable} & {\bf Values}& {\bf Experiment}\\\hline
${\cal A}^{t}_{FB}$    &  $0.19 \pm 0.065$
&{\tt D\O~Collaboration}~~\cite{Abazov:2011rq}\\
                                            & $0.158 \pm 0.074$
& {\tt CDF Collaboration}~\cite{Aaltonen:2011kc}\\
& $0.176\pm 0.05$ & Combined\\
\hline

${\cal A}^{t, low}_{FB}$    &  $0.078 \pm 0.048$ 
&{\tt D\O~Collaboration}~~\cite{Abazov:2011rq}\\   
                                              & $-0.022 \pm 0.043$ 
& {\tt CDF Collaboration}~\cite{Aaltonen:2011kc}\\
& $0.023\pm 0.032$ & Combined\\
\hline
${\cal A}^{t, high}_{FB}$ &  $0.115 \pm 0.060$ 
& {\tt D\O~Collaboration}~~\cite{Abazov:2011rq}\\   
                                              &  $0.266 \pm 0.062$ 
& {\tt CDF Collaboration}~\cite{Aaltonen:2011kc}\\
& $0.188\pm 0.043$ & Combined\\
\hline
$\sigma^{Tevatron}_{t\bar{t}}$ &  $8.18^{+0.98}_{-0.87}$ pb 
& {\tt D\O~Collaboration}~~\cite{Abazov:2009ae}\\\hline
$\sigma^{LHC}_{l^\pm l^\pm}$ &  $< 1$ fb &{\tt ATLAS} $\&$ {\tt CMS Collaborations}~\cite{sstlhc} \\ \hline\hline
\end{tabular}
}
\caption{Measured values of various observables used in our analysis; combined here means weighted averages.}
\label{t:obs}
\end{table}

The most general form of all possible sets of operators containing a 
color-neutral or colored scalar, a vector or a 
tensor can be written 
as follows,

\begin{eqnarray}
Q^{V_s}_{AB}&=&\left ( \overline u \gamma^\mu P_A t\right )\left ( \overline u \gamma_\mu P_B t\right ), \nonumber\\
Q^{V_o}_{AB}&=&\left ( \overline u \gamma^\mu P_A T^a t\right )\left ( \overline u \gamma_\mu P_B T^a t\right ), \nonumber\\
Q^{S_s}_{AB} &=& \left ( \overline u  P_A t\right )\left ( \overline u  P_B t\right ),  \nonumber\\
Q^{S_o}_{AB} &=& \left ( \overline u  P_A T^a t\right )\left ( \overline u  P_B T^a t\right ), \nonumber\\
Q^{T_s}_{A}&=&\left ( \overline u \sigma^{\mu\nu} P_A t\right )\left ( \overline u \sigma_{\mu\nu} P_A t\right ), \nonumber\\
Q^{T_o}_{A}&=&\left ( \overline u \sigma^{\mu\nu} P_A T^a t\right )
\left ( \overline u \sigma^{\mu\nu} P_A T^a t\right ),
\label{e:sstopr}
\end{eqnarray}

\noindent where, $P_A = \frac12 (a_1 + a_5 \gamma^5)$, and, $P_B = 
\frac12 (b_1 + b_5 \gamma^5)$. For $P_A = P_L, a_1 = 1 = - a_5$ so 
for our model we keep $a_{1, 5} ~{\rm and}~ b_{1, 5}$ arbitrary so that 
we can study their dependence as well.

If we consider the case where $AB\in\{LL,RR,LR\}$ and $A\in\{L,R\}$ then 
there are 16 operators in this basis. Due to Fermi statistics of the 
identical t- and u-quarks these operators are not all linearly 
independent. Such a set consisting of 8 independent operators can be 
written as follows:

\begin{eqnarray}
Q_1&=&-\frac13 Q^{S_s}_{LL}+Q^{S_o}_{LL}    \nonumber\\
Q_2&=&\frac23 Q^{S_s}_{LL}+Q^{S_o}_{LL}    \nonumber\\
Q_3&=&-\frac13 Q^{S_s}_{RR}+Q^{S_o}_{RR}    \nonumber\\
Q_4&=&\frac23 Q^{S_s}_{RR}+Q^{S_o}_{RR}    \nonumber\\
Q_5&=&-\frac13 Q^{S_s}_{RL}+Q^{S_o}_{RL}    \nonumber\\
Q_6&=&\frac23 Q^{S_s}_{RL}+Q^{S_o}_{RL}    \nonumber\\
Q_7&=&\frac23 Q^{V_s}_{LL}+Q^{V_o}_{LL}    \nonumber\\
Q_8&=&\frac23 Q^{V_s}_{RR}+Q^{V_o}_{RR}  
\label{Qi_defined}
\end{eqnarray}

\noindent In~\cite{celine} a further condition is imposed that the 
operators are components of operators
respecting the 
$SU(2)\times U(1)$ 
symmetry of the SM. These operators are linear 
combinations of $Q_{5-8}$ so $Q_{1-4}$ are eliminated if one imposes 
this further condition.
In particular the uutt component 
of operators from that paper can be written in terms of the above as 
follows:

\begin{eqnarray}
O_{RR}=Q_8 \ \ \ \ O^{(1)}_{LL}=Q_7\ \ \ \ O^{(3)}_{LL}=Q_7
\nonumber\\
O^{(1)}_{LR}=Q_6-Q_5\ \ \ \  O^{(8)}_{LR}=\frac13 Q_6-\frac23 Q_5
\end{eqnarray}

Let us now consider the dimension 6 effective Lagrangian for $q\overline 
q\to t\overline t$ ($q=u,d$). Generically there are the 20 
possible operators for each choice of $q$ which we may list according to 
a similar scheme as Eqn.~\ref{e:sstopr}:

\begin{eqnarray}
^qS^{V_i}_{AB}
&=&
\left ( \overline q^a \gamma^\mu P_A t^b\right )
\left ( \overline t^c \gamma_\mu P_B q^d\right )
,  \nonumber\\
^qS^{S_i}_{AB}
&=&
\left ( \overline q^a  P_A t^b\right )
\left ( \overline t^c  P_B q^d\right )
,  \nonumber\\
^qS^{T_i}_{A}
&=&
\left ( \overline q^a \sigma^{\mu\nu} P_A t^b\right )
\left ( \overline t^c \sigma_{\mu\nu} P_A q^d\right )
\label{e:ostopr}
\end{eqnarray}

\noindent
here $q\in\{u,d\}$ and $a,b,c,d$ are color indices where 
$i\in\{s,o\}$ indicates whether the color structure is singlet 
($\delta^{ab}\delta^{cd}$) or octet ($T_m^{ab}T_m^{cd}$).

The list constructed in this way contains some redundancy because some 
of the operators are self-conjugate while others appear in conjugate 
pairs.  Enumerating the distinct operators we obtain the following:

\begin{eqnarray}
^qS_1^i&=& ^qS^{S_i}_{LR}      \nonumber\\
^qS_2^i&=& ^qS^{S_i}_{RL}      \nonumber\\
^qS_3^i&=& ^qS^{V_i}_{LL}      \nonumber\\
^qS_4^i&=& ^qS^{V_i}_{RR}      \nonumber\\
^qS_5^i&=& ^qS^{V_i}_{LR}      \ \ \ \ \ 
^qS_5^{i\dag}= {^qS^{V_i}_{RL}}      \nonumber\\
^qS_6^i&=& ^qS^{S_i}_{LL}      \ \ \ \ \ 
^qS_6^{i\dag}= {^qS^{S_i}_{RR}}      \nonumber\\
^qS_7^i&=& ^qS^{T_i}_{L}      \ \ \ \ \ 
^qS_7^{i\dag}= {^qS^{T_i}_{R}}      
\label{s_operators}
\end{eqnarray}

\noindent Thus, operators $^qS^i_{1-4}$ are self-conjugate and so must 
have real coefficients. The operators $^qS^i_{5-7}$; $^qS^{i\dag}_{5-7}$ 
are conjugate pairs. The coefficients of conjugate pairs must be 
conjugate; if they are complex then the model violates CP.

Again, we can identify the operators which are consistent with $SU(2)$. 
Denoting by $u_R$, $d_R$ and $t_R$ the right handed components of 
the respective quarks and by $q_L=\left ( \begin{array}{c} u_L\\ 
d_L\end{array} \right )$ and $Q_L=\left ( \begin{array}{c} t_L\\ 
b_L\end{array} \right )$ the left handed light and heavy doublets 
respectively, these operators are:

\begin{eqnarray}
T_1^i &=& (\overline u_R \gamma^\mu t_R)(\overline t_R \gamma_\mu u_R) 
={^uS_4}
\nonumber\\
T_2^i &=& (\overline u_R  Q_L)(\overline Q_L  u_R) 
={^uS_1}+\dots
\nonumber\\
T_3^i &=& (\overline u_R \gamma^\mu t_R)(\overline Q_L \gamma_\mu q_L) 
={^uS_4}+\dots
\nonumber\\
T_4^i &=& (\overline q_L \gamma^\mu Q_L)(\overline Q_L \gamma_\mu q_L) 
={^uS_3}+{^dS_3}+\dots
\nonumber\\
T_5^i &=& (\overline q_L \gamma^\mu \tau^i Q_L)
(\overline Q_L \gamma_\mu \tau^i q_L) 
={^uS_3}-{^dS_3}+\dots
\nonumber\\
T_6^i &=& (\overline q_L \gamma^\mu  t_R)(\overline t_R \gamma_\mu  q_L) 
={^uS_2}+{^dS_2}
\nonumber\\
T_7^i &=& (\overline d_R \gamma^\mu t_R)(\overline t_R \gamma_\mu d_R) 
={^dS_4}
\nonumber\\
T_8^i &=& (\overline d_R  Q_L)(\overline Q_L  d_R) 
={^dS_1}+\dots
\nonumber\\
T_9^i &=& (\overline d_R \gamma^\mu t_R)(\overline Q_L \gamma_\mu d_L) 
={^dS_4}+\dots
\end{eqnarray}

\noindent The ellipses indicates additional terms which do not 
contribute to $\overline qq\to \overline tt$. Thus only operators 
$^qS_{1-5}^i$ are consistent with $SU(2)$ while $^qS_{6-7}^i$ are not. 
The only potentially CP violating operator consistent with $SU(2)$ is 
therefore $^qS_{5}^i$.

\section{\label{sst:corr}Correlation between the $A^t_{FB}$ and the $\sigma_{tt}$}

One key feature of analyzing new physics contribution to a process in 
terms of an effective Lagrangian is that as the mass scale becomes 
large, only the lowest dimension terms will contribute. This offers the 
prospect of reducing all possible models to a finite number of 
coefficients of operators which can, in principle, be associated with 
different signals. For this to work it is important to quantify how 
small the scale of new physics can be for this parameterization to be 
accurate.

\begin{figure}
\centerline{\includegraphics[angle=0, width=0.65\textwidth]{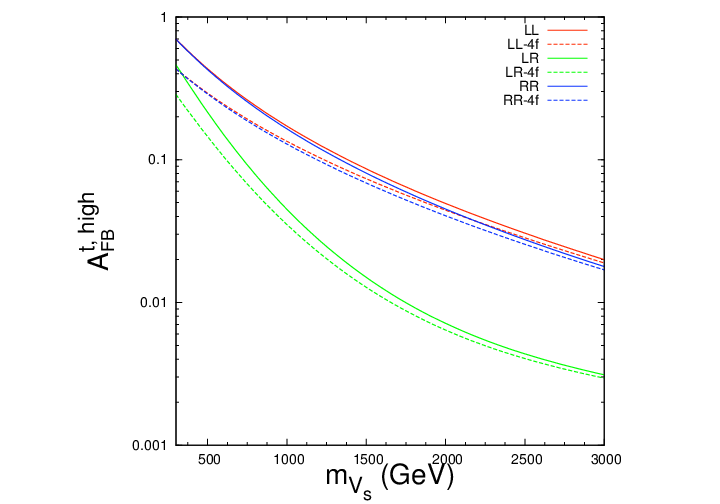}}
\centerline{\includegraphics[angle=0, width=0.65\textwidth]{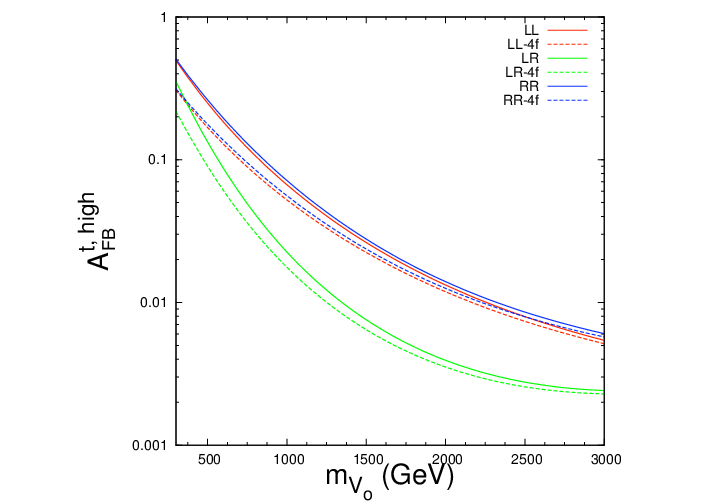}}
\caption{\sf\small 
Tevatron high mass forward-backward asymmetry, ${\cal A}^{t, high}_{FB}$, vs $M_X$ for various cases. In all 
cases, the solid lines show the exchange including a t-channel form 
factor while the dashed lines show the dimension four fermi operator 
without such a form factor. The upper graphs shows cases which are color 
singlet while the lower case shows color octet. In both graphs the red 
lines indicate LL vector couplings, the blue lines indicate RR vector 
couplings and the green line indicates LR vector couplings.
In all cases $g_X=1$.
}
\label{f:v_afb_mx}
\end{figure}

In Figure~\ref{f:v_afb_mx} we address this issue by comparing the 
t-channel exchange of either a massive scalar or vector particle, $X$, 
with the dimension 6 approximation to that interaction.

\begin{figure}
\centerline{
\includegraphics[angle=0, width=0.5\textwidth]{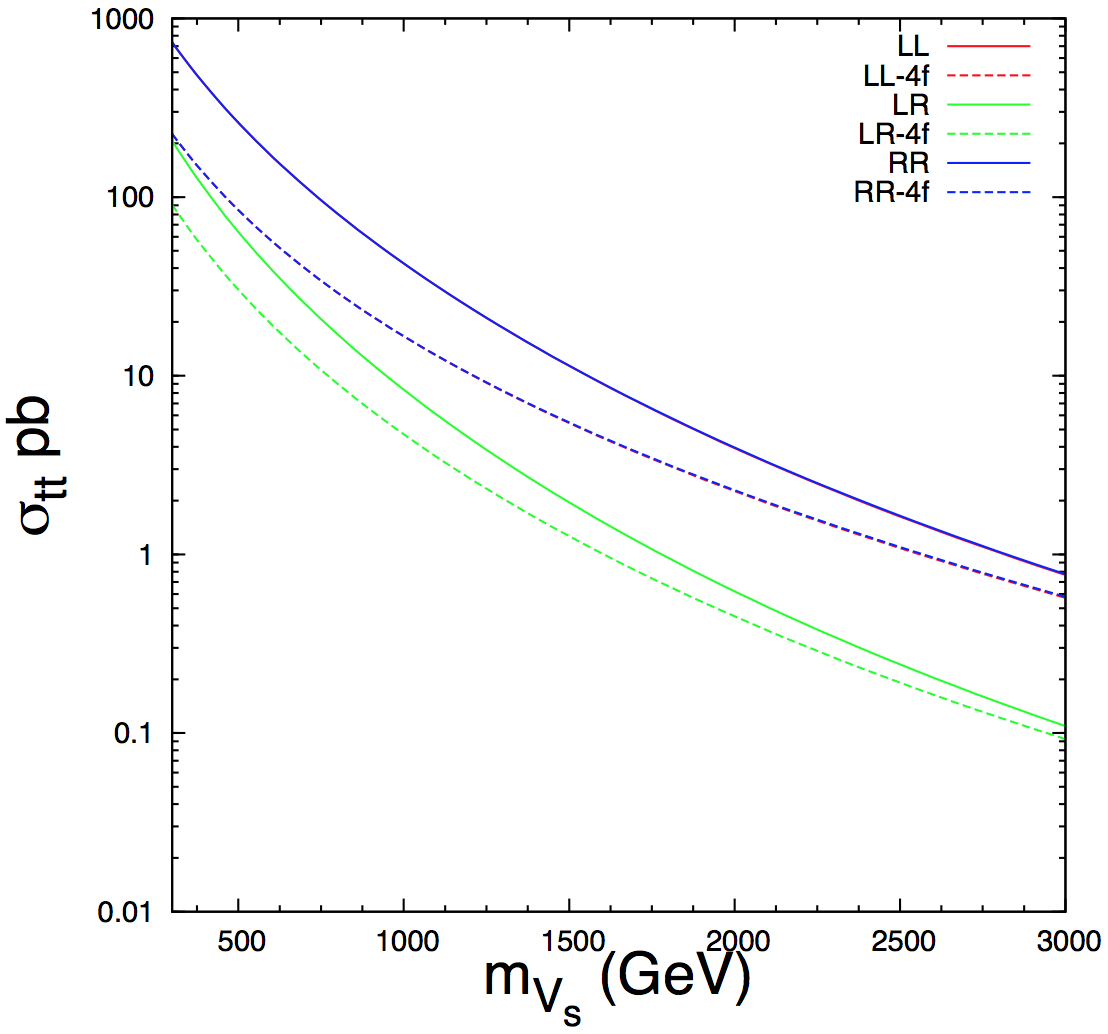}
}
\caption{\sf\small 
The same-sign top pair cross section at the LHC with $\sqrt{s} = 7$ TeV for 
for various color singlet vector cases. 
In all 
cases, the solid lines show the exchange including a t-channel form 
factor while the dashed lines show the dimension four fermi operator 
without such a form factor. 
The red 
lines indicate LL vector couplings, the blue lines indicate RR vector 
couplings and the green line indicates LR vector couplings.
In all cases $g_X=1$. Note that the LL and RR curves fall on top of each 
other since they are related by parity. 
}
\label{f:v_tt_mx}
\end{figure}

In Figure~\ref{f:v_tt_mx} we apply such a comparison to the case of same 
sign top production. In the solid lines we consider a model where the 
amplitude is equal to

\begin{eqnarray}
\frac{g^2}{t-M_X^2}
\overline u \Gamma_1 t\ \ 
\overline u \Gamma_2 t
\end{eqnarray}

\noindent where $\Gamma_i$ (which may contain Lorentz or color indices) 
determines the nature of the exchanged entity. As indicated several 
choices for $\Gamma_i$ are considered in this figure.

The dimension 6 model for such a general particle exchange is recovered 
by setting $t\to 0$ in the above expression. The cross section for this 
result is given by the dashed line. In both cases we have set $g_X = 1$ for 
the normalization for the cross sections which are taken at 
$\sqrt{s} = 7$ TeV and include the branching ratio of tops decaying 
leptonicaly.  In addition we have applied the basic cuts that the 
transverse momentum of a lepton or a jet is $P_T > 25$ GeV while the rapidity 
of these objects is $|\eta| < 2.7$ and their separation is $\Delta R > 0.4$. 
A missing transverse energy cut of ${\slashed E}_T > 30$ GeV is also applied. 
The dimension 6 limit appears to agree with the model within $\lsim 30\%$ when the mass of the exchange particle $M_X\geq 2$ TeV. Below that 
point the two models begin to diverge. Note that since there is only one 
amplitude this conclusion is independent of the assumed value of $g_X$.

In Figure~\ref{f:v_afb_mx} we carry out the same exercise for the 
contribution of various t-channel exchanged objects to the 
forward-backward asymmetry of top-quark production. In this case the 
dimension 6 model agrees to a good approximation ($\lsim 20\%$) with the exchange model when 
$M_X > 1$ TeV.

For cases which could provide an explanation for the observed value of 
${\cal A}_{FB}$ a more general class of models must be considered which 
includes operators of dimension higher than 6. This will necessarily 
constrain somewhat the generality of the conclusions we can reach.

In the cases we are considering, all the higher dimensional operators which 
can contribute at tree level are derived from the dimension 6 set by the 
insertion of pairs of covariant derivatives. In momentum space this 
means that the Feynman rules for these operators are modified by a form 
factor which depends on the kinematics of the scattering. Without loss 
of generallity we can write such a form factor as a function of $t$ and 
$u$ where in the reaction $u(p_1) u(p_2)\to t(p_3) t(p_4)$ or $u(p_1) 
\overline u(p_2)\to t(p_3) \overline t(p_4)$ we define $t=(p_1-p_3)^2$ 
and $u=(p_1-p_4)^2$.

If we generalize the operators in Eqn.~\ref{Qi_defined} we need to take 
into account the correct symmetry under the exchange of identical 
fields. Thus for each operator we introduce two form factors 
$q^+_i(t,u)$ and $q^-_i(t,u)$ which have the symmetries: 
$q^+_i(t,u)=q^+_i(u,t)$ and $q^-_i(t,u)=-q^-_i(u,t)$. In terms of these 
symmetric and ant-symmetric functions we define:

\begin{eqnarray}
q^s(t,u)&=&\frac23q^+_i(t,u)-\frac13q^-_i(t,u)
\nonumber\\
q^{s\prime}(t,u)&=&-\frac13q^+_i(t,u)+\frac23q^-_i(t,u)
\nonumber\\
q^o(t,u)&=&\frac{1}{2}\left(q^+_i(t,u)+q^-_i(t,u)\right)
\end{eqnarray}

The generalization of these operators is thus:

\begin{eqnarray}
\hat Q_1&=&
q^{s\prime}_1(t,u) Q^{S_s}_{LL}+q^{o}_1(t,u)  Q^{S_o}_{LL} 
\nonumber\\
\hat Q_2&=&
q^{s}_2(t,u) Q^{S_s}_{LL}+q^{o}_2(t,u)  Q^{S_o}_{LL}
\nonumber\\
\hat Q_3&=&
q^{s\prime}_3(t,u) Q^{S_s}_{RR}+q^{o}_3(t,u)  Q^{S_o}_{RR}  
\nonumber\\
\hat Q_4&=&
q^{s}_4(t,u) Q^{S_s}_{RR}+q^{o}_4(t,u)  Q^{S_o}_{RR}
\nonumber\\
\hat Q_5&=&
q_5^{s\prime}(t,u)Q^{S_s}_{RL}    
+q_5^{o}(t,u)Q^{S_o}_{RL}    
\nonumber\\
\hat Q_6&=&
q^{s}_6(t,u) Q^{S_s}_{RL}+q^{o}_6(t,u)  Q^{S_o}_{RL}
\nonumber\\
\hat Q_7&=&
q^{s}_7(t,u) Q^{V_s}_{LL}+q^{o}_7(t,u)  Q^{V_o}_{LL}
\nonumber\\
\hat Q_8&=&
q^{s}_8(t,u) Q^{V_s}_{RR}+q^{o}_8(t,u)  Q^{V_o}_{RR}
\label{hatQi_defined}
\end{eqnarray}

Likewise in the case of $q\overline q\to t\overline t$ we can generalize 
the operators given in Eqn.~\ref{s_operators} by replacing their 
coefficients in the effective lagrangian with a form factor depending on 
s and t. In this case, there is no constraint arising from the fermi 
symmetry of identical particles.  We denote the form factors which 
apply to the operator $^qS^i_j$ (see eq. 5) as $^qs^i_j(s,t)$.

Models which are well approximated by the dimension 6 Lagrangian 
correspond to cases where the form factors are constant. Without 
considering fully general form factors, we consider the class of models 
where both $uu\to tt$ and $u\overline u \to t \overline t$ are generated 
by a the exchange of a single species of particle, $X$, in the t-channel 
of mass $M_X$ where $X$ may be either a scalar or a vector and also could 
be either a color singlet or an octet.

In general the form factor for a 
$uu\to tt$ process would be:

\begin{eqnarray}
q_{i}^\pm &=& C_{ut}^2\left ( \frac{1}{t - M_X^2}\pm \frac{1}{u - M_X^2}\right)
\nonumber
\end{eqnarray}

\noindent
where the choice of $i$ depends on the spin and helicity structure of the coupling of $X$ to the quarks and $C_{ab}$ is a matrix in flavor space of the couplings of $X$ to quarks. Likewise in the case of $u\overline u \to t\overline t$ the form factor will be

\begin{eqnarray}
^us^i_j &=& C_{ut}C_{tu}\left ( \frac{1}{t-M_X^2}\right)
\nonumber
\end{eqnarray}

\noindent
We will further assume that the coupling matrix is self-conjugate, in particular, it has the symmetry that 
$|C_{ut}|=|C_{tu}|$. Therefore we define 
$|C_{ut}|^2=g_X^2$. 

\begin{figure}
\centerline{\includegraphics[angle=0, width=1\textwidth]{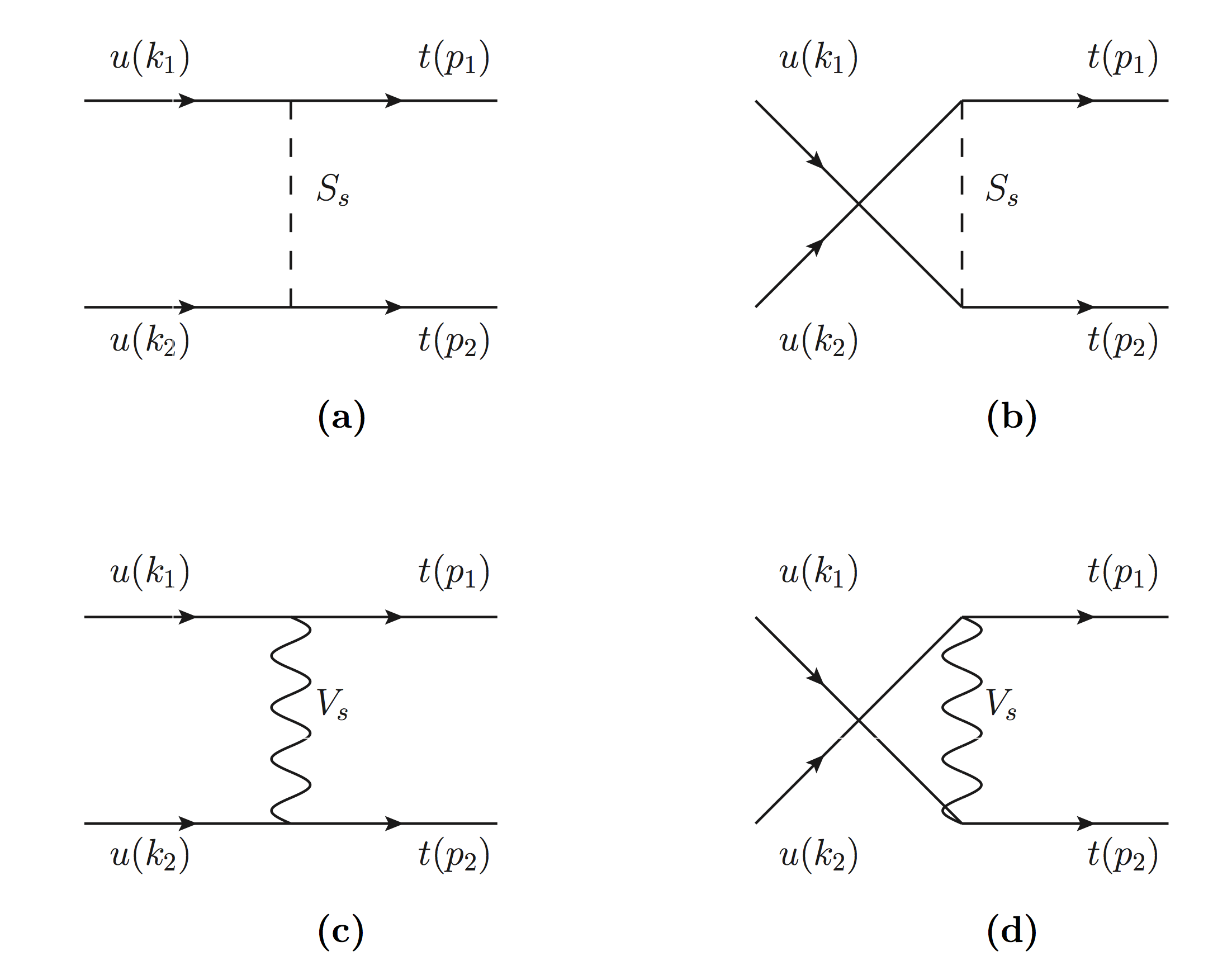}}
\caption{\sf\small Representative parton level Feynman diagrams for the same-sign top pair production at the LHC via a scalar-singlet ($S_s$) exchange (Figs. (a) and (b)), and, a vector-singlet 
($V_s$) exchange (Figs. (c) and (d)).}

\label{f:fg_sst}
\end{figure}

With these assumptions then, a given type of $X$ will generate form factors for both the processes $uu\to tt$ (see, Fig.~\ref{f:fg_sst}) and $u\overline u\to t\overline t$.
In the limit of large $M_X$ this model will correspond to the dimension 6 
Lagrangian with coefficients $-g_X^2/M_X^2$.

For our purposes we considered the following six realistic possibilities 
to each of these with a coupling constant $g_X$;

\begin{itemize}
\item $P_A = P_L = P_B$,
\item $P_A = P_L, P_B = P_R$ or vice-versa,
\item $P_A = P_R = P_B$,
\item $P_A = I = P_B$,
\item $P_A = I, P_B = \gamma_5$ or vice-versa, and, 
\item $P_A = \gamma_5 = P_B$.
\end{itemize}

%

%

%

%
\begin{figure}
\centerline{\includegraphics[angle=0, width=0.6\textwidth]{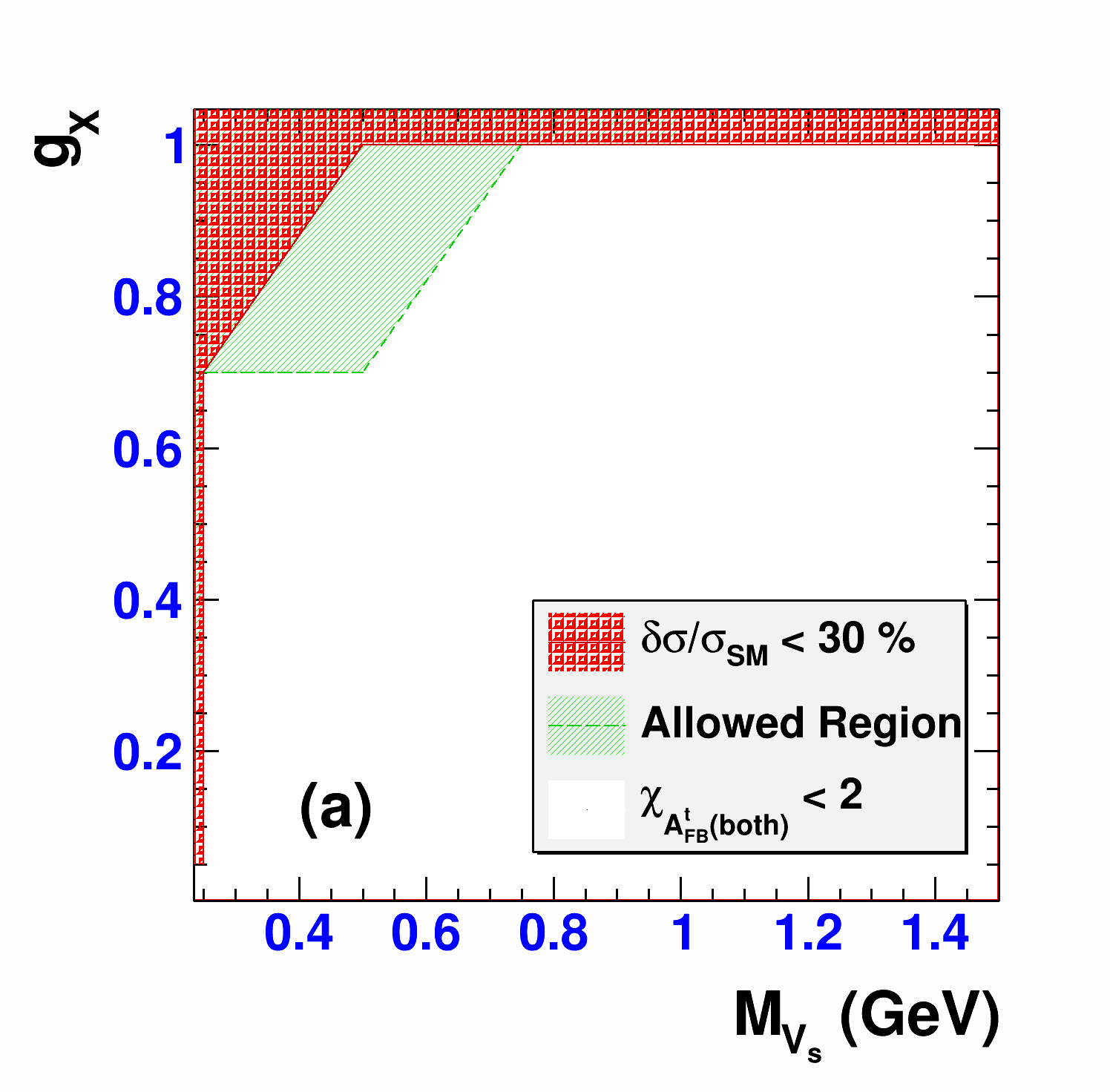}}
\centerline{\includegraphics[angle=0, width=0.6\textwidth]{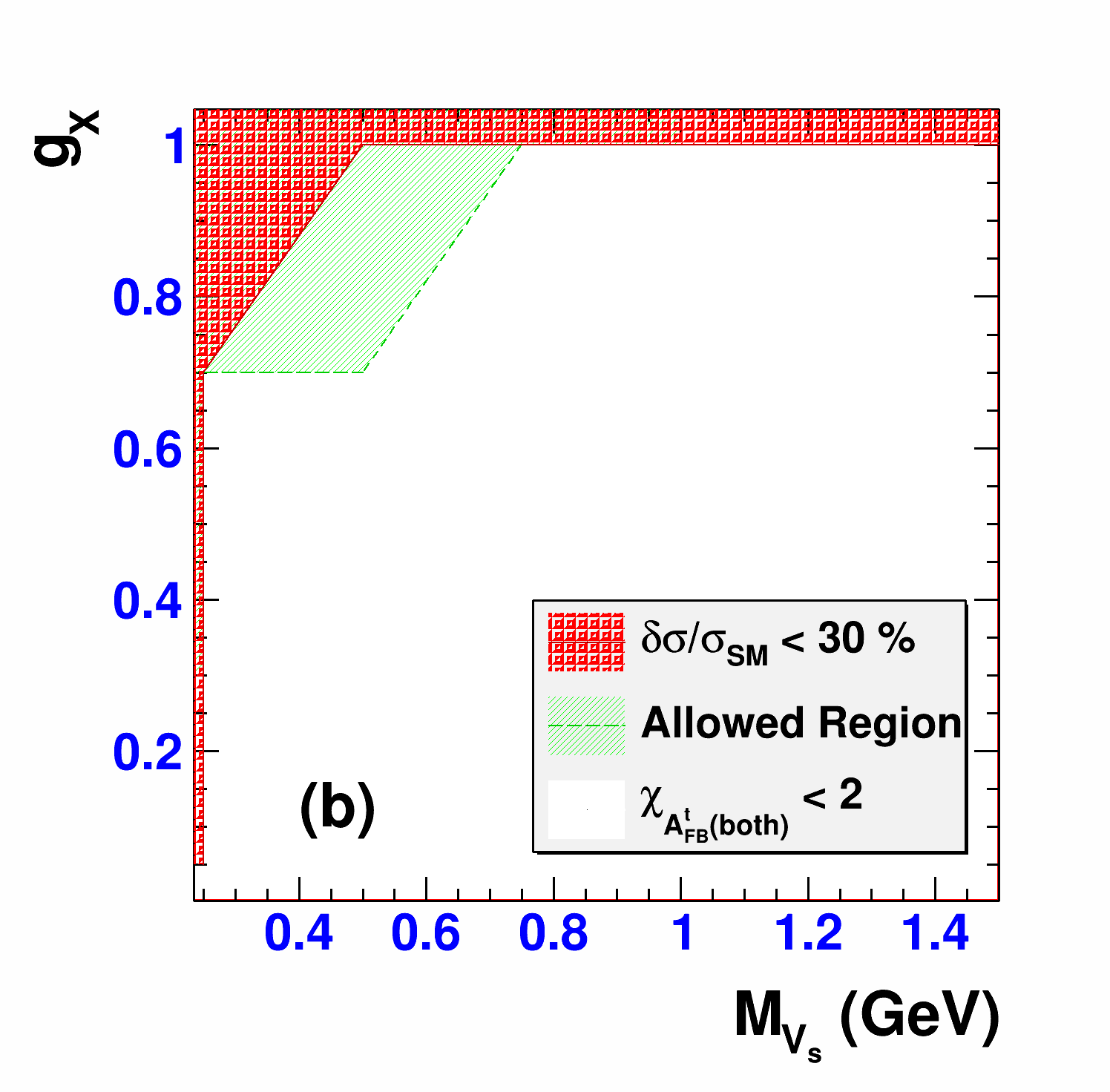}}
\caption{\sf\small 
The allowed parameter space from the existing $t\overline t$ data in the 
$M$ versus $g_X$ plane is shown for the color singlet vector case. The 
couplings combinations considered are (a) $11$, and (b) $55$ (see Eqns. 1 for detail). The 
constraints imposed are that the low and the high mass integrated asymmetries, ${\cal 
A}^{t, low}_{FB}$ ad ${\cal A}^{t, high}_{FB}$ remain within the $2\sigma$ limits (symbolically labelled by ${\mathlarger \chi}_{A^t_{FB}{~\rm (both)} } < 2$) from the experimental 
measurements (indicated by the region below the green dashed line) and 
the new physics contribution to the $t\bar{t}$ cross-section at the 
Teavtron is within $30\%$ of the SM $t\bar{t}$ (the square hatched 
region above the red solid line). The diagonal relation between these two 
lines is thus the region which is allowed by these constraints.
}
\label{f:v_allowed}
\end{figure}
\begin{figure}
\centerline{\includegraphics[angle=0, width=0.6\textwidth]{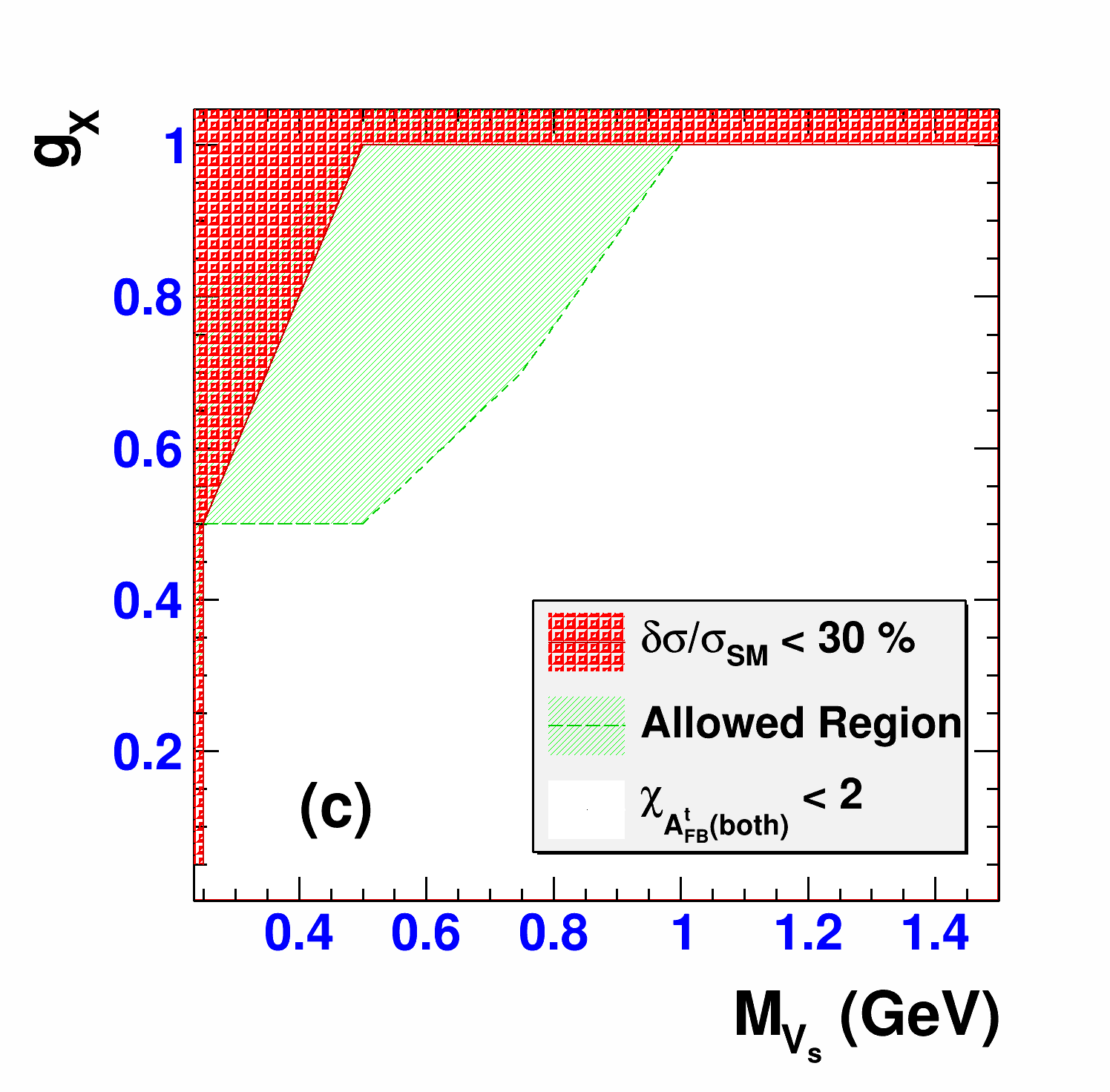}}
\centerline{\includegraphics[angle=0, width=0.6\textwidth]{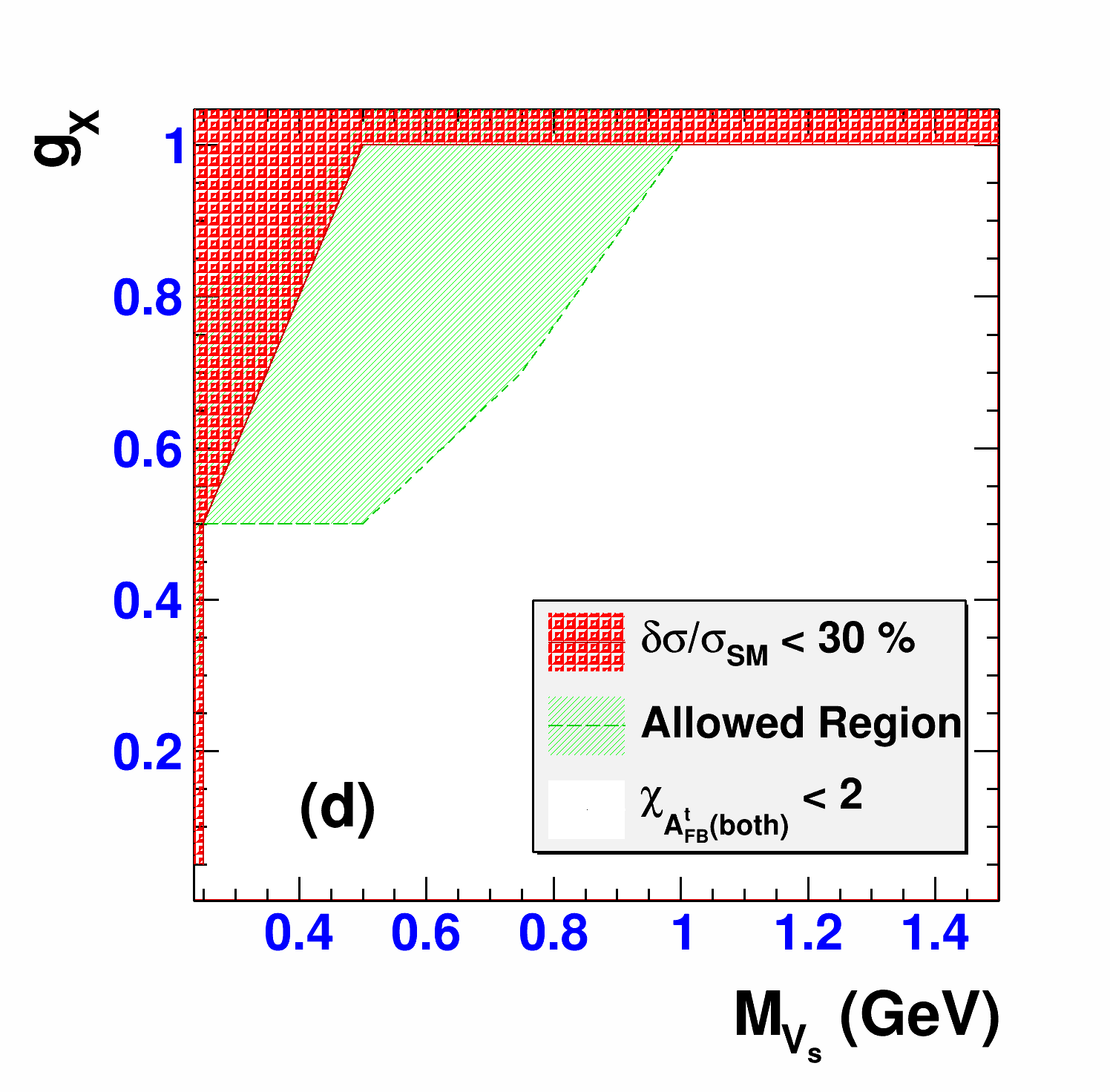}}
\caption{\sf\small 
The exclusion plots here are similar to those of Figure~\ref{f:v_allowed}
except that the vector couplings are (c) LL and (d) RR.
}
\label{f:v_allowed_B}
\end{figure}


This thus leads to $6\times6 = 36$ independent cases for our analysis. 
Apart from the SM parameters, every case will have the mass of the 
exchanged particle $M_X$ and the flavor-violating coupling constant 
$g_X$.

The SM and NP contributions to the top-quark forward-backward 
asymmetries ${\cal A}^{t, SM}_{FB}$ and ${\cal A}^{t, NP}_{FB}$ are 
correlated to the $t\bar{t}$ cross-section in the following manner

\begin{eqnarray}
{\cal A}^{t, Total}_{FB} &=& 
\left(\frac{\sigma^{SM}_{t\bar{t}}}{\sigma^{Total}_{t\bar{t}}}\right) 
{\cal A}^{t, SM}_{FB} +\left(\frac{\sigma^{NP}_{t\bar{t}}}
{\sigma^{Total}_{t\bar{t}}} \right) {\cal A}^{t, NP}_{FB}
, \nonumber \\
{\cal A}^{t, SM}_{FB} &=& \frac{\sigma^{SM}_{t\bar{t}} 
(\cos\theta > 0) - \sigma^{SM}_{t\bar{t}} (\cos\theta < 0)}
{\sigma^{SM}_{t\bar{t}}}, \nonumber \\
{\cal A}^{t, NP}_{FB} &=& \frac{\sigma^{NP}_{t\bar{t}} 
(\cos\theta > 0) - \sigma^{NP}_{t\bar{t}} (\cos\theta < 0)}
{\sigma^{NP}_{t\bar{t}}};
\end{eqnarray}

\noindent where $\sigma^{Total}_{t\bar{t}} = \sigma^{SM}_{t\bar{t}} + 
\sigma^{NP}_{t\bar{t}}$, $\theta$ being the angle of the top quark in 
the $t\bar{t}$ centre-of-mass frame. Note also that the superscript NP (for new physics) here means any specific beyond the SM operator. 

\section{\label{sst:anal}Numerical Analysis and Results}
To gain further insight, in our work, rather than concentrating on the total 
${\cal A}^t_{FB}$ we will analyse the integrated ${\cal A}^t_{FB}$ in 
the low top-pair invariant mass region ($< 450$ GeV) and in the high top-pair 
invariant mass region ($\geq 450$ GeV), i.e. ${\cal A}^{t, low}_{FB}$ and ${\cal A}^{t, high}_{FB}$ respectively. The measurements of each of 
these are listed in Table~\ref{t:obs}. Thus at least the following four 
observables need to be satisfied by any operator that aims to provide an 
experimentally viable solution to the ${\cal A}^t_{FB}$, 

(a) top-pair production cross-section, $\sigma_{t\bar{t}}$, 

(b) the integrated asymmetry in low $m_{t\bar{t}}$ ($ < 450$ GeV) region, ${\cal A}^{t, low}_{FB}$,

(c) the integrated asymmetry in high $m_{t\bar{t}}$ ($ \geq 450$ 
GeV) region, and, ${\cal A}^{t, high}_{FB}$, and,

(d) same-sign top pair cross-section at the LHC, 
$\sigma_{tt}$.

For our numerical analysis we implemented the aforementioned model into 
the package {\tt MadGraph5}~\cite{madgraph} using {\tt 
FeynRules}~\cite{feynrules} and reproduced some of the results in Ref.~\cite{celine} for checks. To evaluate the parton densities we use 
{\tt CTEQ6L1}~\cite{cteq}. The renormalisation scale $\mu_R$, and the 
factorisation scale, $\mu_F$ are fixed at $\mu_R = \sqrt{\hat{s}} = 
\mu_F$. We scan the following ranges of the $M_X-g_X$ parameter space: 
$g_X$ between $10^{-2}$ to $1$ and $M_X$ between $0.2$ TeV to $4$ TeV.

We analyse each operator individually in the following three steps: 
In order to find the parameter space favoured by various operators as 
mentioned in Eqns.~\ref{e:sstopr}, we first ensure that for each of the 
allowed parameter space point the new physics contribution to the 
$t\bar{t}$ remains below $30\%$ of the SM $t\bar{t}$ cross-section,  
To implement the constraints due to low and high mass region integrated 
asymmetries we first combine their measured values by the {\tt CDF} and 
{\tt D\O} into corresponding weighted averages as mentioned in Table 1 and then allow only those parameter space points for which the theoretical asymmetries due to each of the two $m_{t\bar{t}}$ regions are 
within $2\sigma$ standard deviation from the measured values. \\ Finally we enforce that the same-sign leptons (which 
arise due to the leptonic decays of the produced pair of same-sign tops) 
cross-section also remains within the LHC non-observation limits~\cite{sstlhc} for the 
current data on such events.

In order to demonstrate the dependence of ${\cal A}^t_{FB}$ in 
Figs.~\ref{f:v_afb_mx} we plot ${\cal A}^{t, high}_{FB}$ for the color 
singlet and octet vector cases with $LL$, $LR$ and $RR$ operators for a 
fixed values of $g_X = 1$. Clearly a wide $M_X$ range below $\lsim 2$ 
TeV is capable of producing the desired asymmetry for LL and RR singlet 
vector cases. This range becomes relatively narrower ($\lsim 1$ TeV) for 
the mixed case with LR couplings. The cases with octet vectors are more 
restricted (i.e. only $M_X \sim$ several hunders of GeVs are allowed) because their contribution to the ${\cal A}^t_{FB}$ is 
suppressed by the color-factor of $3/8$.

\begin{table}[]
\centering
{
\begin{tabular}{ | l | l | l | l| l|}\hline\hline
\emph{\bf Operator}& \multicolumn{4}{|c|}{\bf Observables}\\
\cline{1-5}
 & \emph{$ \frac{\delta\sigma_{t\bar{t}}}{\sigma^{SM}_{t\bar{t}}} $} 
& \emph{$ + \chi_{_{A^{t, low}_{FB}}}$} 
& \emph{$ + \chi_{_{A^{t, high}_{FB}}} $}  
& \emph{$ + \sigma_{l^\pm l^\pm}$}\\\hline
 \cline{2-5}
\multicolumn{5}{|c|}{\bf Case 1: Singlet Scalar}\\\hline\hline
$Q^{S_s}_{LL}$ & \yes & \yes & \yes & \nil \\\hline
$Q^{S_s}_{LR}$ & \yes & \yes & \nil & \nil \\\hline
$Q^{S_s}_{RR}$ & \yes & \yes & \yes & \nil \\\hline
$Q^{S_s}_{11}$ & \yes & \yes & \nil & \nil \\\hline
$Q^{S_s}_{15}$ & \yes & \yes & \nil & \nil \\\hline
$Q^{S_s}_{55} $ & \yes & \yes & \nil & \nil \\\hline
\hline
 \cline{2-5}
\multicolumn{5}{|c|}{\bf Case 2: Octet Scalar}\\\hline\hline
$Q^{S_o}_{LL}$ & \yes & \yes & \nil & \nil \\\hline
$Q^{S_o}_{LR} $ & \yes & \yes & \nil & \nil \\\hline
$Q^{S_o}_{RR} $ & \yes & \yes & \nil & \nil \\\hline
$Q^{S_o}_{11} $ & \yes & \yes & \nil & \nil \\\hline
$Q^{S_o}_{15} $ & \yes & \yes & \nil & \nil \\\hline
$Q^{S_o}_{55}$ & \yes & \yes & \nil & \nil \\\hline
\hline
 \cline{2-5}
\multicolumn{5}{|c|}{\bf Case 3: Singlet Vector}\\\hline\hline
$Q^{V_s}_{LL} $                  & \yes & \yes & \yes & \nil \\\hline
$Q^{V_s}_{LR} $                  & \yes & \yes & \nil & \nil \\\hline
$Q^{V_s}_{RR} $                  & \yes & \yes & \yes & \nil\\\hline
$Q^{V_s}_{11} $                      & \yes & \yes & \yes & \nil \\\hline
$Q^{V_s}_{15}$               & \yes & \yes & \nil & \nil \\\hline
$Q^{V_s}_{55}$        & \yes & \yes & \yes & \nil \\\hline
\hline
 \cline{2-5}
\multicolumn{5}{|c|}{\bf Case 4: Octet Vector}\\\hline\hline
$Q^{V_o}_{LL}$           & \yes & \yes & \nil & \nil \\\hline
$Q^{V_o}_{LR}$           & \yes & \yes & \nil & \nil \\\hline
$Q^{V_o}_{RR}$           & \yes & \yes & \nil & \nil \\\hline
$Q^{V_o}_{11} $               & \yes & \yes & \nil & \nil \\\hline
$Q^{V_o}_{15}$        & \yes & \yes & \nil & \nil \\\hline
$Q^{V_o}_{55}$ & \yes & \yes & \nil & \nil \\\hline
\hline
\end{tabular}
}
\caption{\sf \small 
Allowed (\yes) and Disallowed (\nil) operators for singlet and octet 
scalar ($S_s, S_o$), and, vector ($V_s, V_o$) cases for various 
observables. The allowed region should satisfy the following constraints 
on the Tevatron and LHC observables: $(1) \left| 
\frac{\delta\sigma_{t\bar{t}}}{\sigma^{SM}_{t\bar{t}}} \right | < 0.3$, 
where $\delta\sigma_{t\bar{t}}= \sigma^{Total}_{t\bar{t}} - 
\sigma^{SM}_{t\bar{t}}$ measured difference in the $t\bar{t}$ 
cross-section at the Teavtron, $(2)~ \chi_{_{A^{t, low}_{FB}}} < 2$, 
$(3)~ \chi_{_{A^{t, high}_{FB}}} < 2$ that means the $\chi^2 < 2$ for the asymmetry in the low and in the high mass region from experiment; and, $(4)~ \sigma_{l^\pm l^\pm} < 
1$ fb with $l^\pm = e^\pm$ and $\mu^\pm$, $\sigma_{l^\pm l^\pm} $ is the 
same-sign dilepton cross-section for the LHC at $\sqrt{s} = 7$ TeV. The 
analysed ranges for exchanged particle mass, and the couplings are: $M_X 
\in \left[0.25, 4\right]$ TeV, and $g_X \in \left[0.01, 1\right]$.
}
\label{t:analysis}
\end{table}

To demonstrate how strong these individual constraints can be, in 
Fig.~\ref{f:v_allowed} we show some of vector cases where at least some 
of the parameter space is allowed (marked with diagonal lines) after imposing the requirement that 
the new physics contribution to the total $t\bar{t}$ cross-section at 
the Tevatron should not exceed $30 \%$ and also the results must be consistent 
with the measured ${\cal A}^t_{FB}$ at $2\sigma$ level. Thus we find 
that in order to produce sufficiently large ${\cal A}^t_{FB}$ consistent 
with the experiment, $g_X$ must be sufficiently large and/or $M_X$ must 
be small while at the same time smaller correction in 
$\sigma_{t\bar{t}}$ requires the $g_X$ to be small and/or $M_X$ larger 
which essentially means that the two constraints push the $g_X - M_X$ 
from the two extremes.

In Fig.~\ref{f:v_tt_mx} we plot the same-sign top-pair cross-sections 
for the LHC at 7 TeV for fixed values of $g_X = 1$. To this end it is 
worth emphasising that for fixed $M_X$, $\sigma_{tt}(g_X)|_{M_X} = g_X^4 
\times \sigma_{tt}(g_X = 1)|_{M_X}$. Also we expect the efficiency of 
cuts to be the same for fixed $M_X$. Clearly an order of magnitude 
suppression in $g_X$ would mean four orders of reduction in the 
same-sign top-pair event rate for the fixed $M_X$.

The aforementioned production processes give rise to a pair of same-sign 
leptons associated with a b-jet pair when both the produced tops decay 
leptonically in turn. Because so far LHC did not observe such events~\cite{sstlhc}, we 
therefore would like to analyse the above couplings in light of the 
LHC non-observation constraints for this process.

\subsection{SM background}

We now would like to address the SM background that could possibly 
confuse the aforementioned signal. The most serious contribution to SM 
background for this process is due to the production of 
$pp\longrightarrow W^+j W^+j$, where $W^+ \longrightarrow l^+ \nu_l$. 
Using {\tt MadGraph5}, we estimate the bare cross-section for the final 
process $pp \longrightarrow W^+j W^+j \longrightarrow (l^+ \nu_l) j (l^+ 
\nu_l) j$ to be of about $3.66$ fb for the $\sqrt{s} = 7$ TeV at the 
LHC, which reduces to $1.53$ fb when we demand the following basic cuts 
on the leptons and the jets: $p_{T_{l, j}} > 25$~GeV, $\left|\eta_{_{l, 
j}}\right| \leq 2.7$; $\Delta{R}_{l,l},\Delta{R}_{l,j} \geq 0.4$ and 
$\not\!\!E_T > 30~{\rm GeV}$.

The above SM background final-state topology can also be obtained when 
two partons from the same proton scatter into a $W^+$ and a jet and a 
the same time partons from the other proton also scattering into another 
$W^+ j$. This is known as Double-parton scattering (DPS) and its 
cross-section can be roughly estimated by~\cite{dps},

\begin{eqnarray}
\sigma^{DPS}(W^+j W^+ j) = \frac{\sigma(W^+j)^2}{ 2 \sigma^{eff.}},
\end{eqnarray}

\noindent where, $\sigma^{eff.}$ is the total effective cross-section at 
the LHC which has been measured to be about $11$ mb according to 
\cite{eff}. Using the above formula, we estimate, $\sigma^{DPS}(W^+j W^+ 
j)$ = 0.9 fb before any cuts which becomes 0.02 fb when we allow both 
the W's leptonically and fold-in the basic cuts as discussed above.

The final contribution to the above background is due to the pile-up at 
the LHC which can be estimated using the following formula,

\begin{eqnarray}
\sigma^{Pile}(W^+j W^+ j) = \frac{\sigma(W^+j)^2}{ 2 \sigma^{inel.}} N_p,
\end{eqnarray}

where $\sigma^{inel.}$ is the total inelastic cross-section at the LHC, 
and, $N_p$ is the number of pile-up events per bunch-crossing which are 
measured to be $110$ mb and 32 respectively.~\cite{pileup}. We therefore 
obtain $\sigma^{Pile}(W^+j W^+ j) = 2.88$ fb without any cuts which 
translates into $0.06$ fb same-sign lepton events. Thus the total 
background cross-section for the same-sign lepton has been estimated to 
be about $1.53 + 0.02 + 0.06 = 1.61$ fb.

Now because LHC did not observe such events, in order to apply the 
constraint due to such a process, we demand signal significance for this 
process to be less than 1 for the 5 fb$^{-1}$ data which is equivalent 
to having less than 8 signal events for the 5 fb$^{-1}$ assuming the 
events are distributed according to Poisson distribution.

For our analysis of same-sign lepton pairs, we first produce the 
same-sign top pairs using {\tt MadGraph} and then allow them to decay 
semi-leptonically into a b-jet and a $\mu^\pm$ or $e^\pm$ via a $W^\pm$. 
We then impose the same basic cuts as mentioned before and finally 
demand 8 or lesser same-sign lepton pair events for $\int {\cal L} dt = 
5 fb^{-1}$ integrated LHC-7 luminosity. We found that for $M_X = 500$ 
GeV, these basic cuts reduce the same-sign lepton event rate by about 
$\epsilon_C =42\%$. Now since each top-quark decays into a lepton about 
$25\%$ of the time times when incorporating decays of leptonic decays of 
taus, therfore the effective event rate for the same-sign lepton pair 
would be $\sigma_{l^\pm l^\pm} = \sigma_{tt} \times {Br(t\to b + l + 
...)}^2 \times \epsilon_C \times \epsilon_b^2 \simeq 0.01 \sigma_{tt}$, where, $\epsilon_b$, the b-tagging efficiency has been assumed to be $\sim 58 \%$ at the LHC~\cite{btag}.
Thus the observed $\sigma_{l^\pm l^\pm}$ is only about $1$ {percent} of 
the raw $\sigma_{tt}$. \\ \\

\begin{table}[]
\centering
{
\begin{tabular}{ | l | l | l | l| l|}\hline\hline
\emph{\bf Operator}& \multicolumn{4}{|c|}{\bf $M_X$ (TeV)}\\
\cline{1-5}
 & \emph{$ 0.25$} & \emph{$ 0.5$} & \emph{$1 $}  & \emph{$2$}\\\hline
 \cline{2-5}
\multicolumn{5}{|c|}{\bf Case 1: Singlet Scalar}\\\hline\hline
$Q^{S_s}_{LL/RR}$ & 0.1 & 0.3& 0.5 & 1 \\\hline
$Q^{S_s}_{LR}$ & 0.3&  0.5& 0.7 & 1 \\\hline
$Q^{S_s}_{11/55}$ & 0.3 & 0.5 & 0.7 & 1 \\\hline
$Q^{S_s}_{15}$ & 0.3 & 0.5 & 0.9 & 1 \\\hline
\hline
 \cline{2-5}
\multicolumn{5}{|c|}{\bf Case 2: Octet Scalar}\\\hline\hline
$Q^{S_o}_{LL/RR}$ & 0.3 & 0.5 & 0.8 & 1 \\\hline
$Q^{S_o}_{LR} $ & 0.3 & 0.5 & 0.9 & 1 \\\hline
$Q^{S_o}_{11/55} $ & 0.5 & 0.7 & 0.8 & 1 \\\hline
$Q^{S_o}_{15} $ & 0.7 & 0.9 & 1 & 1 \\\hline
\hline
 \cline{2-5}
\multicolumn{5}{|c|}{\bf Case 3: Singlet Vector}\\\hline\hline
$Q^{V_s}_{LL/RR} $                  & 0.05 & 0.1 & 0.3 & 0.5 \\\hline
$Q^{V_s}_{LR} $                  & 0.1 & 0.1 & 0.3 & 0.7 \\\hline
$Q^{V_s}_{11/55} $                      & 0.1 & 0.1 & 0.3 & 0.7 \\\hline
$Q^{V_s}_{15}$               & 0.1 & 0.3 & 0.4 & 0.7\\\hline
\hline
 \cline{2-5}
\multicolumn{5}{|c|}{\bf Case 4: Octet Vector}\\\hline\hline
$Q^{V_o}_{LL/LL}$           & 0.1 & 0.1& 0.3 & 0.6 \\\hline
$Q^{V_o}_{LR}$              & 0.1 & 0.3& 0.7 & 1 \\\hline
$Q^{V_o}_{11/55} $        & 0.1 & 0.3& 0.7 & 1 \\\hline
$Q^{V_o}_{15}$        & 0.25 & 0.6&  0.7& 1 \\\hline
\hline
\end{tabular}
}
\caption{\sf \small 
Upper bounds on the coupling $g_X$ for various operators after demanding 
that the ${t\bar t}$ cross-section must not deviate from experiment by more than $10 \%$ and the events with same-sign lepton pair are consistent with the SM background 
at the LHC at $\sqrt{s} = 7$ TeV and at integrated luminosity, $\int{\cal L} dt =  5$ fb$^{-1}$.}
\label{t:10pc_sst}
\end{table}

\section{\label{sst:concl}Conclusions}

We conclude our findings in Table~\ref{t:analysis} and ~\ref{t:10pc_sst} 
for all the cases. Table~\ref{t:analysis} shows that although all the 
operators are allowed by the $t\bar{t}$ cross-section data and by ${\cal A}^{t, low}_{FB}$, most of them are ruled out by ${\cal A}^{t, high}_{FB}$ at the $2 \sigma$ (or equivalently $95\%$ 
{\em confidence level}). Out of all the cases with scalar, vector and 
tensor operators, only two cases with a scalar singlet operators, 
$Q^{S_s}_{LL}$ and $Q^{S_s}_{RR}$, four cases with a singlet vector 
operators, $Q^{V_s}_{LL}$, $Q^{V_s}_{RR}$, $Q^{V_s}_{11}$, and 
$Q^{V_s}_{55}$ are not ruled out by the Tevatron data. Later these also get excluded by the imposition of the LHC constraint on the $\sigma_{l^\pm 
l^\pm}$. We are thus left with no allowed case by all four observables 
we analysed. This suggests that it is extremely difficult if not 
impossible, to provide a tree-level solution to the ${\cal A}^t_{FB}$ 
on the basis of pure (self-conjugate) flavor-changing top interactions.

Finally we also checked the viability of each of the operators when 
imposing only the total $t\overline t$ cross section and the same sign 
top cross-section, i.e. suppose we drop the ${\cal A}^t_{FB}$ constraints 
altogether assuming that ${\cal A}^t_{FB}$ is not a genuine effect due to new physics\footnote{We thank Sally Dawson for this nice idea.}. For this 
purpose we assume that the top-cross-section measurement has only $10 \%$ errors. Table~\ref{t:10pc_sst} summarizes the resulting 
upper bounds on the $g_X$ for a wide range of $M_X$ for various 
operators. Clearly this suggest that bounds due to non-observation of 
same-sign leptons at the LHC are not as strong as the one which arise 
due to the imposition of ${\cal A}_{FB}$ constraints.

Thus, using an extremely general phenomenological Lagrangian we have 
shown that all models with t-channel self-conjugate flavor-changing resonance up to a scale of 4 
TeV, that have been proposed to account for ${\cal A}^t_{FB}$, are ruled out. Since tree-level flavor-changing operators seem unable to account for the forward-backward
top-quark production asymmetries and stay simultaneously consistent with the same-sign top-quark production data,  if the experimental deviations  of the forward-backward asymmetry from the Standard Model
gets firmly established,  one may need to resort to flavor-changing loop level new physics
interactions; an example of this is studied in~\cite{DMS_12}.

\section*{Acknowledgements}
We thank Alex Kagan for a useful communication. The work of D.~A. and A.~S. are supported in part by US DOE grant Nos. DE-FG02-94ER40817 (ISU) and 
DE-AC02-98CH10886 (BNL). The work of S.~K.~G. was supported in part by 
the {\em ARC Centre of Excellence for Particle Physics at the 
Tera-scale}. The use of Monash University Sun Grid, a high-performance 
computing facility, is gratefully acknowledged.

\end{document}